\documentclass[fleqn,usenatbib]{mnras}

\usepackage{newtxtext,newtxmath}

\usepackage{subcaption}
\usepackage[T1]{fontenc}

\DeclareRobustCommand{\VAN}[3]{#2}
\let\VANthebibliography\thebibliography
\def\thebibliography{\DeclareRobustCommand{\VAN}[3]{##3}\VANthebibliography}

\usepackage{graphicx}	
\usepackage{amsmath}	

\title[MOOSE III]{Monitoring Observations of SMC X-1's Excursions (MOOSE) III: X-ray Spectroscopy of a Warped, Precessing Accretion Disc}

\author[R. Karam et al.]{
Rawan Karam,$^{1,2,3}$\thanks{E-mail: rewan.karam@mail.mcgill.ca} Kristen C. Dage$^{2,3,4,7}$ \thanks{NASA Einstein Fellow}, Bailey E. Tetarenko,$^{2,3}$ McKinley C. Brumback$^{5,6}$, Daryl Haggard$^{2,3}$,\newauthor
 Arash Bahramian,$^{7}$, Chin-Ping Hu,$^{8}$ Joey Neilsen, $^{9}$ 
Diego Altamirano,$^{10}$
Wasundara Athukoralalage,$^{11,12}$
\newauthor
Philip A. Charles, $^{10}$
William I. Clarkson,$^{13}$
Ryan C. Hickox,$^{14}$
Jamie Kennea$^{15}$
\\
$^{1}$ Département de Physique, Université de Montréal, Succ. Centre-Ville, Montréal, Québec, H3C 3J7, Canada \\
$^{2}$Department of Physics, McGill University, 3600 University Street, Montr\'eal, QC H3A 2T8, Canada\\
$^{3}$Trottier Space Institute at McGill 3550 University Street, Montr\'eal, QC H3A 2A7, Canada \\
$^{4}$ Wayne State University, Department of Physics \& Astronomy, 666 W Hancock St,  Detroit, MI 48201, USA \\
$^{5}$Department of Physics, Middlebury College, Middlebury, VT 05753, USA\\
$^{6}$Department of Astronomy, University of Michigan, 1085 S.\ University Ave. Ann Arbor, MI 48109 USA\\
$^{7}$International Centre for Radio Astronomy Research $-$ Curtin University, GPO Box U1987, Perth, WA 6845, Australia\\
$^{8}$ Department of Physics, National Changhua University of Education, Changhua, 50007, Taiwan \\
$^{9}$ Villanova University, Department of Physics, Villanova, PA 19085, USA \\
$^{10}$Physics \& Astronomy, University of Southampton, Southampton, Hampshire SO17 1BJ, UK\\
$^{11}$ Center for Astrophysics | Harvard \& Smithsonian, 60 Garden Street, Cambridge, MA 02138-1516, USA \\
$^{12}$Department  of  Physics  and  Astronomy,  Michigan  State  University,  East Lansing, MI 48824, USA\\
$^{13}$Department of Natural Sciences, University of Michigan-Dearborn, 4901 Evergreen Rd. Dearborn, MI 48128, USA\\
$^{14}$ Department of Physics \& Astronomy, Dartmouth College, 6127 Wilder Laboratory, Hanover, NH 03755, USA\\
$^{15}$ Department of Astronomy and Astrophysics, The Pennsylvania State University, University Park, PA 16802, USA }

\date{Accepted XXX. Received YYY; in original form ZZZ}

\pubyear{2024}

\newcommand{\nh} {$N_{\text{H}}$}

\begin{document}
\label{firstpage}
\pagerange{\pageref{firstpage}--\pageref{lastpage}}
\maketitle

\begin{abstract}

The MOOSE (Monitoring Observations of SMC X-1’s Excursions) program uses the Neutron Star Interior Composition Explorer Mission (NICER) to monitor the high mass X-ray binary SMC X-1 during its superorbital period excursions. Here we perform X-ray spectral analyses of 26 NICER observations of SMC X-1, taken at the tail-end of the excursion between 2021-04-01 and 2022-01-05. We use a single spectral model to fit spectra observed in high, intermediate and low states, using a combination of a partial covering fraction model, a black-body disc, and a power-law component. We find that the partial covering fraction varies significantly with the superorbital state during superorbital excursion. Our findings suggest that the low/high state in SMC X-1 is caused by a very high obscuration of the accretion disk. 

\end{abstract}




\section{Introduction}

Warped accretion discs challenge our current models and understanding of accretion physics. Many studies now suggest that warped accretion discs are far more ubiquitous than previously thought, and are present in black holes and neutron star systems alike \citep{Townsend20,Thomas2022}. For example, \citet{Clarkson03} demonstrated that many systems display superorbital periods (detected periods longer than the binary orbit) with varying lengths, which are due to warped accretion discs, where the stability of the warp is determined by properties of the binary system, such as binary separation and binary mass ratio \citep{Ogilvie01}. Smoothed particle hydrodynamic simulations (e.g. \citealt{Foulkes06}) have indicated that the central engine's illumination of the accretion disc can exert non-axisymmetric radiation, which causes tilting or warping on the disc's surface.

\begin{figure*}
    \centering
    \includegraphics[width=6.5in]{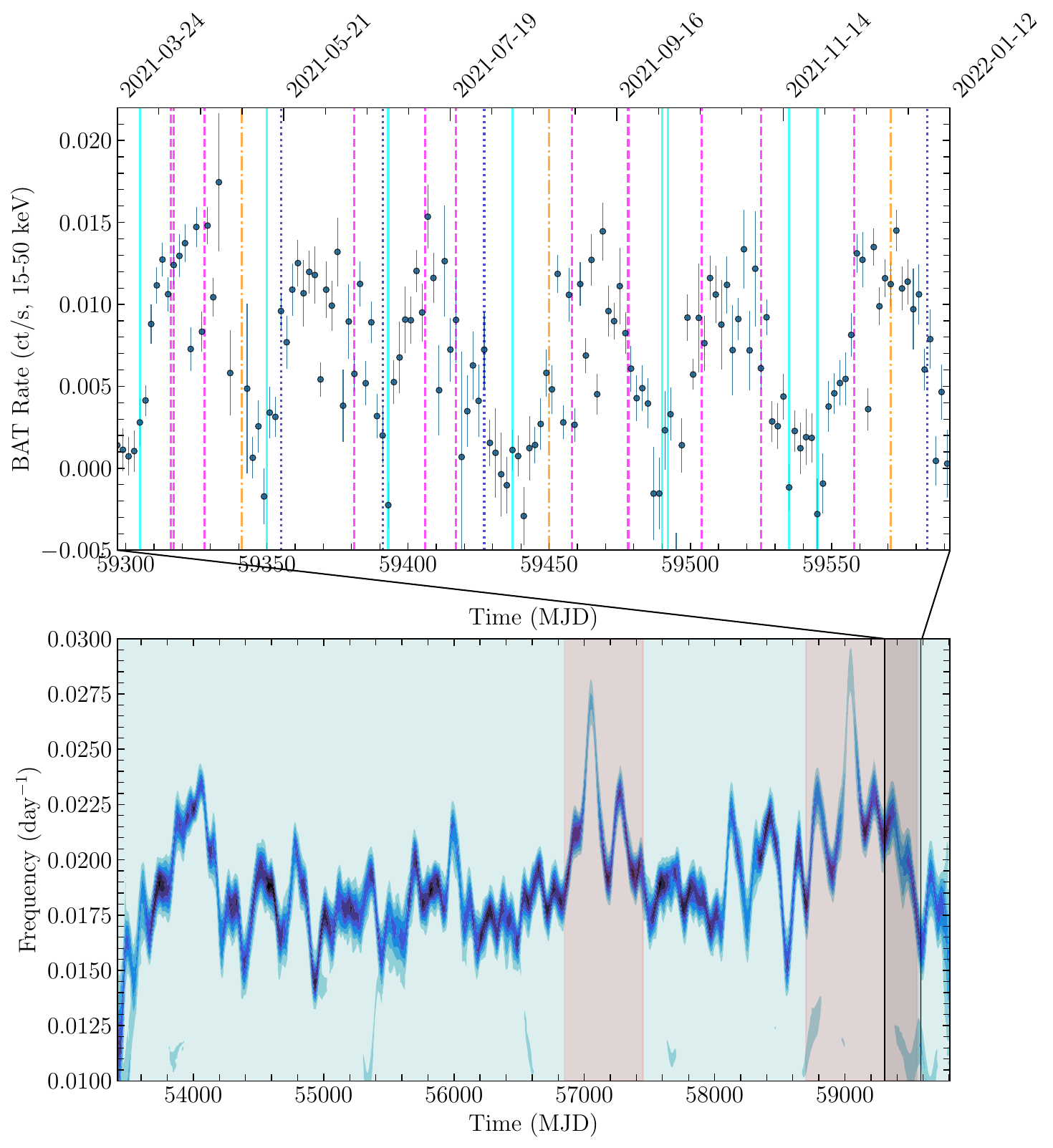}
    \caption{Hilbert-Huang transform of SMC X-1 (from \citealt{Hu23}) showing  three of its superorbital period excursions. The inset at MJD 59305 shows the \textit{Swift}/BAT lightcurve of SMC X-1 overlaid with observation times from our NICER monitoring campaign (pink dashed are high state observations, blue dotted lines are intermediate state observations, aqua solid lines are low state observations and gold dash-dot lines are pre-eclipse dips), taken during the tail-end of the superorbital period excursion. Our observations span a range of superorbital phases and source luminosity. The grey shaded regions mark the duration of the third and fourth superorbital excursions, respectively. Our observations span the tail end of the fourth excursion and continue after the excursion has finished. }  
    \label{fig:observations}
\end{figure*}

The neutron star X-ray binary SMC X-1 is one of the best systems in which to study the physics behind accretion discs due to its well-determined distance, low extinction and variety of time dependent behavior. SMC X-1's orbital period is 3.89 days with a pulse period of 0.7 seconds \citep{luck76}, and it exhibits a superorbital period of 40--60 days \citep{Wojdowski98,Gruber84}. This is likely due to a warped, precessing accretion disc \citep{Clarkson03, Hickox04,Brumback20}. SMC X-1's superorbital period is nominally $\sim$ 55 days, but is observed to change to as low as 45 days (Figure \ref{fig:observations}), a so-called superorbital period ``excursion'', which occurs when the collimated radiation hits the accretion disc, changing its geometry \citep{Foulkes06}. The broad-band X-ray spectrum has been characterized as a cut-off power law continuum with a low energy blackbody component and iron emission lines by various instruments and studies including ASCA \citep{Paul02}, Beppo-SAX \citep{Naik04}, XMM-Newton and NuSTAR \citep{Brumback20}, and NICER \citep{Dage22}.

SMC X-1's superorbital period has been a long-standing subject of investigation since \citet{Gruber84}, which observed the variability and spectrum of SMC X-1 with observations from the UCSD/MIT instrument on HEAO 1 from 1977 to 1979. Their results suggest that SMC X-1 has two states, high or low, and they constructed average spectra separately for each state, which showed that the total emission from SMC X-1 displays a continuum spectrum with a dominant exponential form. More recent studies using XMM-Newton and NuSTAR's combined broad-band X-ray coverage reveal that the superorbital high and low states exhibit different X-ray continua and suggest higher absorption during the low states \citep{Pike19, Brumback20}, consistent with observation from a precessing warped accretion disk. \citet{Pradhan20} examined archival Suzaku and NuSTAR spectra of SMC X-1 during different superorbital states and noted large changes in the normalization of the power law component that suggest an additional mechanism may be driving SMC X-1's spectral changes with superorbital phase.

SMC X-1 also exhibits timing behaviors beyond its superorbital period, including an energy dependent pulse profile and binary eclipses. Beppo-SAX observations of SMC X-1 revealed a single peaked pulse profile at energies below 1 keV and a double peaked profile at higher energies \citep{Naik04}. These results were confirmed by observations using Chandra, XMM-Newton, and NuSTAR. \citet{Hickox05} and \citet{Brumback20} performed pulse-phase resolved spectroscopy of SMC X-1 and found that the blackbody component of the X-ray spectrum followed the low energy pulse profile and the power-law component followed the high energy power law, similar to ASCA studies by \citet{Paul02}. These studies suggest that the soft component is likely produced by reprocessing of the hard X-ray pulsar beam by the inner accretion disc. They support this claim by using a model of a twisted inner disc illuminated by the rotating X-ray pulsar beam to simulate pulsations in the soft component, thus showing that the precession of an illuminated accretion disc can roughly reproduce the observed long-term changes in the soft-pulse profiles for some disc and beam geometries \citep{Hickox05, Brumback20}. 

Studies of the X-ray orbital light curve of SMC X-1 have revealed it to have pre-eclipse dips in the orbital phase range of 0.6-0.85 \citep[e.g.,][]{Woo95,2010MNRAS.401.1532R, Hu13,Brumback22}. While not well studied, these dips are thought to be similar to those seen in Her X-1 \citep{Giacconi73}, and are caused by increased obscuration from the impact of the accretion stream onto the accretion disc.

SMC X-1's superorbital excursions offer a rare chance to study both pulsar behaviour and the accretion disc at the same time. \citet{Dage19} and \citet{Hu19} highlighted that the superorbital period and the spin of SMC X-1 may be both correlated with the physics behind the accretion. \citet{Hu2013} applied the Hilbert-Huang transform to analyse the time-frequency properties of the superorbital modulation in SMC X-1 from observations made by the All-Sky Monitor onboard the Rossi X-ray Timing Explorer.  The resultant Hilbert spectrum showed that the superorbital modulation period varied between $\sim$40 and $\sim$60 days, providing a robust timing technique that could very closely follow the changes in the superorbital period.

Later analysis of SMC X-1 by \citet{Hu19} showed that a third observed superorbital period excursion event occurred during 2014-2016 and suggests that this excursion is recurrent and possibly periodic. \citet{Brumback20} modeled the geometry of SMC X-1 and showed that superorbital period cycles are indeed consistent with a warped accretion disc. However, a similar analysis has not yet been completed for data taken during a superorbital period excursion, and it is not known whether the geometry of the disc changes as the superorbital period changes. Stray-light studies of SMC X-1 with NuSTAR found that SMC X-1's pulse profile did not vary with energy, but did change significantly over time \citep{Brumback22}. 

A study of SMC X-1 by \citet{Pike19} suggests that its transient pulsations may be due to obscuration of the warped accretion disc. Thus, understanding the physical processes behind accretion is the first step to addressing the superorbital period, and understanding SMC X-1 may help shed light on the more extreme processes happening in the more distant pulsating ultraluminous X-ray sources, some of which show similar long-term variability \citep{Bachetti2020, Townsend20}.

The first paper in this series, MOOSE I, \citet{Dage22} introduces the first 26 observations of the MOOSE campaign and provides spectral fits to the high state spectra of SMC X-1 as it exits its fourth epoch of superorbital period excursion (Figure \ref{fig:observations}). In these high-state spectral fits, \citet{Dage22} found very little fluctuation between key spectral parameters like absorption column density, black-body disc temperature, and photon index. MOOSE II \citep{Hu23} studies the excursion phenomenon in SMC X-1 specifically, comparing the properties of the most recent excursion event (2020-2021 or MJD 58700--59550, see Figure \ref{fig:observations}) to prior excursions. Their study of the superorbital period finds that the spin-up acceleration and the pulse profiles (the shape of the pulsations emanating from the neutron star) may be connected to the superorbital excursion. These behaviours suggest that SMC X-1 is a complex physical system that requires more than a simple warped disc model \citep[see also][]{Pradhan20}.

\cite{Dage22} studied 11 NICER high-state observations (out of the 26 observations taken during the tail-end of excursion) and found no evidence for significant changes in the spectral shape of the superorbital high state. In this work, we present the first spectral fits to all 26 of the NICER observations to investigate spectral changes between superorbital high, intermediate, and low states and to examine if the spectral shapes change after the end of excursion. We present the data and analysis in Section \ref{sec:obs} and discuss the results in Section \ref{sec:res}. Our conclusions and recommendations for future work are presented in Section \ref{sec:conc}.

\begin{table*}
    
    \caption{NICER observations log with count rate errors determined by \textsc{xspec}. Pre-eclipse dips are marked with $\dagger$.}
    \label{table:observations}
\begin{tabular}{l|c|c|c|c|c}
\hline
ObsID & Date & Duration (s) & Overonly Range & Underonly Range & Count Rate (ct/s)  \\ 
\hline
\hline
4509010101 & 2021-04-01 & 1006 & $<$11  & \textless{}200 & 33 $\pm$  0.2\\
4509010102 & 2021-04-12 & 1362 & $<$3   & \textless{}200 & 214 $\pm$  0.4 \\
4509010103 & 2021-04-13 & 483  & $<$7   & \textless{}200 & 220 $\pm$  0.6\\
4509010301 & 2021-04-24 & 436  & $<$20  & \textless{}200 & 195 $\pm$  0.7 \\
4509010401 $\dagger$ & 2021-05-07 & 967  & $<$40  & \textless{}200 & 2 $\pm$ 0.05\\
4509010501 & 2021-05-16 & 1446 & $<$140 & \textless{}200 & 7 $\pm$  0.07 \\
4509010601 & 2021-05-21 & 1024 & $<$4   & \textless{}200 & 146 $\pm$ 0.4  \\
4509010701 & 2021-06-16 & 1279 & $<$2   & \textless{}200 & 206 $\pm$ 0.4 \\
4509010801 & 2021-06-26 & 1684 & $<$1.5 & \textless{}200 & 139 $\pm$ 0.3 \\
4509010802 & 2021-06-28 & 796  & $<$20  & \textless{}200 & 10 $\pm$  0.2 \\
4509010803 & 2021-07-11 & 944  & $<$30  & \textless{}200 & 232 $\pm$ 0.5 \\
4509010901 & 2021-07-22 & 821          & $<$20           & \textless{}200  & 179 $\pm$  0.4 \\
4509011001 & 2021-08-01 & 1235         & $<$50           & \textless{}200  & 72 $\pm$  0.3 \\
4509011101 & 2021-08-11 & 1123         & $<$1            & \textless{}200  & 9 $\pm$ 0.1\\
4509011201 $\dagger$ & 2021-08-24 & 1327         & $<$20           & \textless{}200  & 52 $\pm$ 0.3\\
4509011301 & 2021-09-01 & 1898         & $<$0.5          & \textless{}200  & 189 $\pm$ 0.4\\
4509011401 & 2021-09-21 & 1476         & $<$35           & \textless{}200  & 245 $\pm$ 0.5\\
4509011501 & 2021-10-03 & 350          & $<$16           & \textless{}200  & 9 $\pm$ 0.2\\
4509011601 & 2021-10-05 & 1107         & $<$10           & \textless{}200  & 10 $\pm$ 0.1\\
4509011701 & 2021-10-17 & 2533         & \textless{}1   & \textless{}200  & 195 $\pm$ 0.3\\
4509011901 & 2021-11-07 & 1259         & $<$30           & \textless{}200  & 176 $\pm$ 0.4\\
4509012001 & 2021-11-17 & 955          & $<$75           & \textless{}200  & 10 $\pm$ 0.2 \\
4509012101 & 2021-11-27 & 1513         & $<$100          & \textless{}200  & 6 $\pm$ 0.07\\
4509012201 & 2021-12-10 & 1476         & $<$11           & \textless{}200  & 249 $\pm$ 0.4 \\
4509012301 $\dagger$ & 2021-12-23 & 2411         & \textless{}1   & \textless{}306  &  5 $\pm$ 0.06\\
4509012401 & 2022-01-05 & 1474 & \textless{}15 &  \textless{}300& 151 $\pm$ 0.3\\
\hline
\end{tabular}
\end{table*}

\begin{figure*}
    \centering
    \includegraphics[width=6.5in]{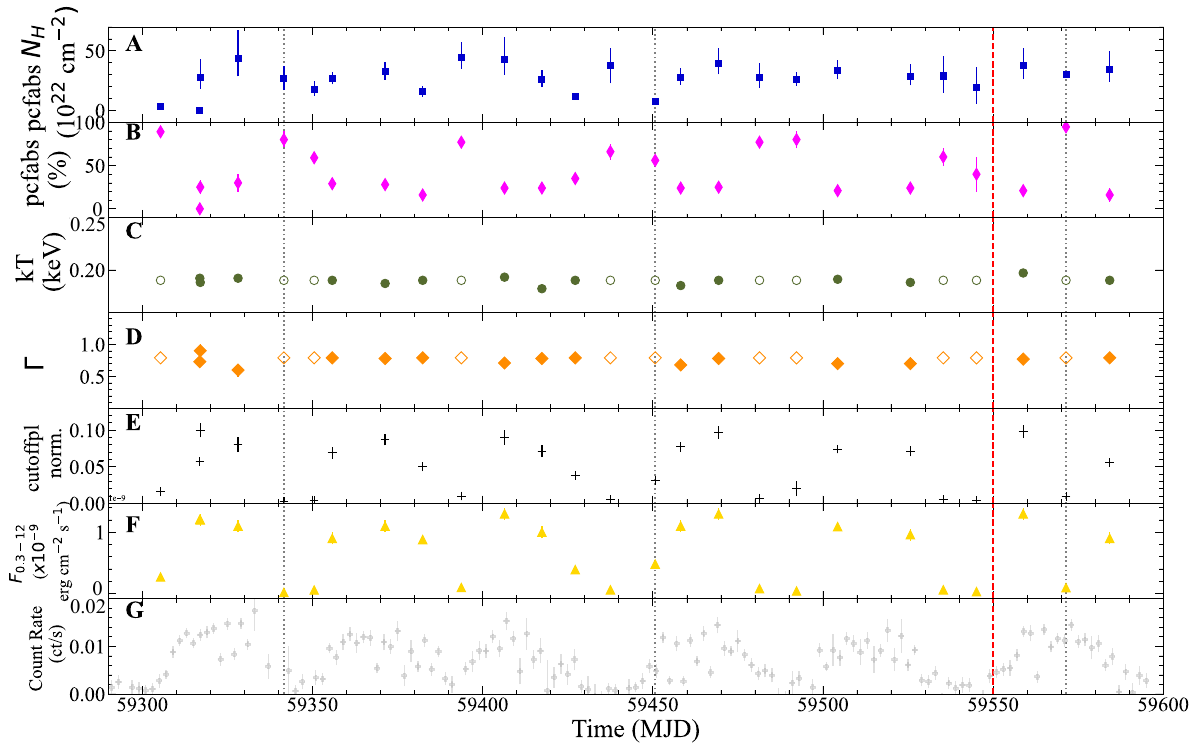}
    \caption{
Variation with time (MJD) of the key SMC X-1 X-ray spectral fit parameters: Panel A (blue points) shows \texttt{pcfabs} $N_H$. Panel B (pink points) shows partial covering fraction. Panel C (green points) shows disc black-body kT. Due to varying spectral quality between high and low state observations, we fix the kT in the low and intermediate state spectra to the average high state value of 0.19. The fixed values are plotted with open symbols, while the free values are plotted with filled symbols. Panel D (orange points) shows the $\Gamma$ values. As in Panel C, reduced \textbf{S/N} in low state prompted us to fix this value in the low state spectra to the average high state value of 0.79. As before, the fixed points are plotted with open symbols while those where $\Gamma$ was left variable are plotted with filled symbols. The impact of fixing these kT and $\Gamma$ values is discussed in \S\ref{sec:xray_spec}. Panel E (black crosses) shows the normalization of the cut-off power law. Panel F (gold triangles) shows the total flux in the NICER band of 0.3-12 keV. Finally, Panel G (grey points) shows the Swift/BAT light curve during the epoch where these observations were taken. The blackbody kT and $\Gamma$ values fluctuate very little with superorbital phase, while the \texttt{pcfabs} $N_H$ and covering fraction both show variations with superorbital period, and in particular the covering fraction shows a mostly smooth variation that is maximized during the superorbital low states. Table \ref{table:fits} details the spectral fit values. In all panels, the red vertical dashed line represents the end of the latest excursion epoch and the grey dotted lines mark the three observations that took place during pre-eclipse dip.}
    \label{fig:allparam}
    
\end{figure*}

\begin{table*}

  \caption{Best fit parameters for our model, \texttt{tbabs*pcfabs*(bbody}+cutoffpl+gauss+gauss+gauss), in the 0.3-12 keV band. Observations in the low state (i.e., with a count rate lower than 50 counts/sec) and intermediate state were fit with a model with blackbody temperature and power law index frozen to the best fit values from the high states (see \S\ref{sec:xray_spec}). The \nh\ for \texttt{tbabs} has been frozen to 2.5 $\times 10^{21} \text{cm}^{-2}$. Pre-eclipse dips are marked with $\dagger$.}
 
\begin{tabular}{l|c|c|c|c|c|c|c|c}
\hline
ObsID & State & \texttt{pcfabs} \nh & kT & \texttt{pcfabs} & $\Gamma$ & Power law norm. & $F_{0.3-12}$  & $\chi^2$ /d.o.f \\
 &  & ($10^{22} ~\text{cm}^{-2}$) & (keV) &  ({\%}) &  &  & (erg cm$^{-2}$ s$^{-1}$)  & \\
\hline
\hline
4509010101 & Low & 3.2 $\pm$ 0.2 & \textit{0.19} & 0.89  $\pm$ 0.02 & \textit{0.79} & 0.0158 $\pm$ 0.0007 & (2.7$\pm$ 0.1)$\times10^{-10}$ & 176/130 \\
\hline
4509010102 & High  &  $< 0.02$  & 0.192 $\pm$ 0.002 & 0.00 $\pm$ 0.05 & 0.73 $^{+0.03}_{-0.05}$& 0.0572$\pm$0.0008  & (1.21$\pm$ 0.02)$\times10^{-9}$ & 300/161 \\
\hline
4509010103 & High & 27.3 $^{+15.6}_{-9.0}$  & 0.191 $\pm$ 0.003 & 0.24  $\pm$ 0.08 & 0.9 $\pm$ 0.1 & 0.10 $\pm$ 0.01 & (1.2$\pm$0.1)$\times10^{-9}$ & 159/147 \\
\hline
4509010301 & High& 43.9 $^{+23.6}_{-15.4}$  & 0.192 $\pm$ 0.004 & 0.3  $\pm$ 0.1 & 0.6 $\pm$ 0.1 & 0.08$\pm$0.01 & (1.1$\pm$0.1)$\times10^{-9}$ & 143/143\\
\hline
4509010401 $\dagger$ & Low& 26.6 $^{+10.8}_{-9.3}$ & \textit{0.19} & 0.8$\pm$0.1 & \textit{0.79} & 0.003$^{+0.005}_{-0.001}$  &  (1.6$\pm$0.2)$\times10^{-11}$ & 139/111 \\
\hline
4509010501 & Low & 17.5 $^{6.9}_{-5.2}$ & \textit{0.19} & 0.59  $\pm$ 0.05 & \textit{0.79} & 0.0036$\pm$0.0006 & (5.6$\pm$0.9)$\times10^{-11}$ & 240/124\\
\hline

4509010601 & Intermediate & 26.6$^{+5.7}_{-4.8}$ & \textit{0.19} & 0.29 $\pm$ 0.07 & \textit{0.79} & 0.069$\pm$0.008 & (9$\pm$1)$\times10^{-10}$ & 177/130 \\
\hline

4509010701 & High&  32.3$^{+8.7}_{-6.7} $ & 0.187 $\pm$ 0.002 & 0.28 $\pm$ 0.05 & 0.78 $\pm$ 0.04 & 0.087$\pm$0.008 & (1.1$\pm$0.1)$\times10^{-9}$ & 268/162 \\
\hline

4509010801 & Intermediate& 15.8 $^{+4.6}_{-4.2}$ & \textit{0.19} & 0.16 $\pm$ 0.05 & \textit{0.79} & 0.050$\pm$0.004 & (8.8$\pm$0.7)$\times10^{-10}$ & 226/160 \\
\hline

4509010802 & Low& 44.8 $^{+12.4}_{-10.0}$ & \textit{0.19}& 0.77 $\pm$ 0.05 & \textit{0.79} & 0.010$\pm$0.002 & (1.0$\pm$0.2)$\times10^{-9}$ & 244/126 \\

\hline

4509010803 &High & 42.8 $^{+18.4}_{-12.9}$  & 0.193 $\pm$ 0.002 & 0.24$\pm$0.07 & 0.71 $\pm$ 0.09  & 0.09$\pm$0.01  & (1.3$\pm$0.1)$\times10^{-9}$ & 187/159 \\

\hline

4509010901 &High & 25.8 $^{+8.1}_{-6.3}$ & 0.182 $\pm$ 0.003 & 0.24  $\pm$ 0.06 & 0.78 $\pm$ 0.09 & 0.071 $\pm$ 0.008 & (1.0$\pm$0.1)$\times10^{-9}$ & 171/156 \\
\hline
4509011001 & Intermediate& 11.3 $\pm$ 2.5 &\textit{0.19}& 0.35  $\pm$ 0.06 & \textit{0.79} &  0.038$\pm$0.005 & (3.9$\pm$0.5)$\times10^{-10}$ & 189/130 \\
\hline
4509011101 & Low& 37.5$\pm$14.8 & \textit{0.19} & 0.66  $\pm$ 0.00 & \textit{0.79} & 0.005 $\pm$ 0.001 & (6$\pm$1)$\times10^{-11}$ & 324/125 \\
\hline
4509011201  $\dagger$ & Low& 7.9 $\pm$ 1.2 & \textit{0.19} & 0.56 $\pm$ 0.02 & \textit{0.79} & 0.031$\pm$0.003  & (4.8$\pm$0.5)$\times10^{-10}$ & 334/130 \\
\hline

4509011301 & High& 27.5$^{+8.1}_{-6.1}$ & 0.185 $\pm$ 0.002 & 0.24  $\pm$ 0.05 & 0.68 $\pm$ 0.08 & 0.077$\pm$0.007 & (1.1$\pm$0.1)$\times10^{-9}$ & 177/160 \\
\hline
4509011401 & High & 39.7$^{+12.7}_{-9.6}$ & 0.190 $\pm$ 0.002 & 0.25  $\pm$ 0.06 & 0.78 $\pm$ 0.07 & 0.097$\pm$0.009 & (1.3$\pm$0.1)$\times10^{-9}$ & 254/161 \\
\hline
4509011501 & Low& 27.9 $^{+12.2}_{-9.0}$ & \textit{0.19} & 0.77  $\pm$ 0.07 & \textit{0.79} & 0.007$\pm$0.005 & (8$\pm$2)$\times10^{-11}$ & 137/117 \\
\hline
4509011601 & Low& 26.2  $^{+6.4}_{-5.6}$ & \textit{0.19} & 0.8  $\pm$ 0.1 & \textit{0.79} & 0.02$\pm$0.01  & (4$\pm$3)$\times10^{-11}$ & 435/130 \\
\hline
4509011701 & High& 33.3 $^{+9.0}_{-7.0}$& 0.191 $\pm$ 0.002 & 0.21  $\pm$ 0.05 & 0.70 $\pm$0.06 & 0.074$\pm$0.005 & (1.09$\pm$0.07)$\times10^{-9}$ & 267/160  \\
\hline
4509011901 & High& 28.6 $^{+10.7}_{-7.6}$ & 0.188 $\pm$ 0.003 & 0.24  $\pm$ 0.06 & 0.7 $\pm$ 0.1 & 0.071$\pm$0.007 & (1.0$\pm$0.1)$\times10^{-9}$ & 170/154 \\
\hline
4509012001 &Low & 28.9  $^{+17.0}_{-14.0}$ & \textit{0.19} & 0.6   $\pm$ 0.1  & \textit{0.79} &  0.005$\pm$0.001 & (6$\pm$1)$\times10^{-11}$ & 282/123 \\
\hline
4509012101 & Low    & 19.2 $^{+17.3}_{-13.8}$   & \textit{0.19} & 0.4  $\pm$ 0.2 & \textit{0.79}  & 0.004$\pm$0.002 & (3$\pm$2)$\times10^{-11}$ & 221/126 \\
\hline
4509012201 & High & 37.6 $^{+15.1}_{-11.3}$  & 0.197 $\pm$ 0.002 & 0.21  $\pm$ 0.06 & 0.77 $\pm$ 0.08 & 0.098$\pm$0.009 & (1.3$\pm$0.1)$\times10^{-9}$ & 223/159 \\
\hline
4509012301  $\dagger$& Low & 30.4 $\pm$ 1.9 & \textit{0.19} & 0.947 $\pm$ 0.004 & \textit{0.79} & 0.0096 $\pm$ 0.0008 & (9.4$\pm$0.8)$\times10^{-11}$ & 441/130 \\
\hline
4509012401 & Intermediate & 34.4 $^{+15.8}_{-10.5}$ &  \textit{0.19} & 0.16 $\pm$ 0.08  & \textit{0.79}  & 0.056$\pm$0.006 & (9$\pm$1)$\times10^{-10}$ & 193/130  \\
\hline
\end{tabular}
   \label{table:fits}
\end{table*}
\section{Observations and Data Analysis}

\label{sec:obs}
We analysed 26 observations within the MOOSE data set of SMC X-1, taken with the Neutron Star Interior Composition Explorer (NICER; \citealt{2016SPIE.9905E..1HG}). These observations were taken in random intervals of at least 10 days, spanning different stages of the superorbital cycle (specifically the high, intermediate and low states), from 2021-04-01 (MJD 59305) until 2022-01-05 (MJD 59584). Most observations were taken towards the end of the 2020-2021 superorbital excursion, defined as the time interval from MJD 58700 to 59550 \citep{Hu23}. The final three observations occurred after this excursion epoch (see Figure \ref{fig:observations} and Table \ref{table:observations}).

\subsection{NICER} 
The NICER observations were processed with \textsc{heasoft} 6.29 \textsc{nicerl2} task. For each observation, the spectrum was generated with the \textsc{xselect} command, with optimal binning implemented (\citealt{Kaastra16}). For the background, the spectra were generated with the \textsc{nibackgen3C50} tool \citep{2022AJ....163..130R}. The \textsc{nicerarf} and \textsc{nicerrmf} tasks were used to generate the ancillary response files (ARF) and response matrix files (RMF). This data reduction follows the method used in \citet{Dage22}, where it is described in detail. Further information including date, exposure, count rate, and other filtering parameters can be found in Table \ref{table:observations}.

\subsection{X-ray Spectral Fitting}
\label{sec:xray_spec}
Working with these spectra, we tested a series of models to fit the 0.3-12 keV energy range for all observations. We used \textsc{xspec} v12.12.0 \citep{Arnaud96} to model the spectra, using $\chi^2$ statistics, abundances from \cite{Wilm}, and photo-electric cross sections from \cite{1996ApJ...465..487V}.

\begin{figure*}
    \centering
    \includegraphics[scale=0.45]{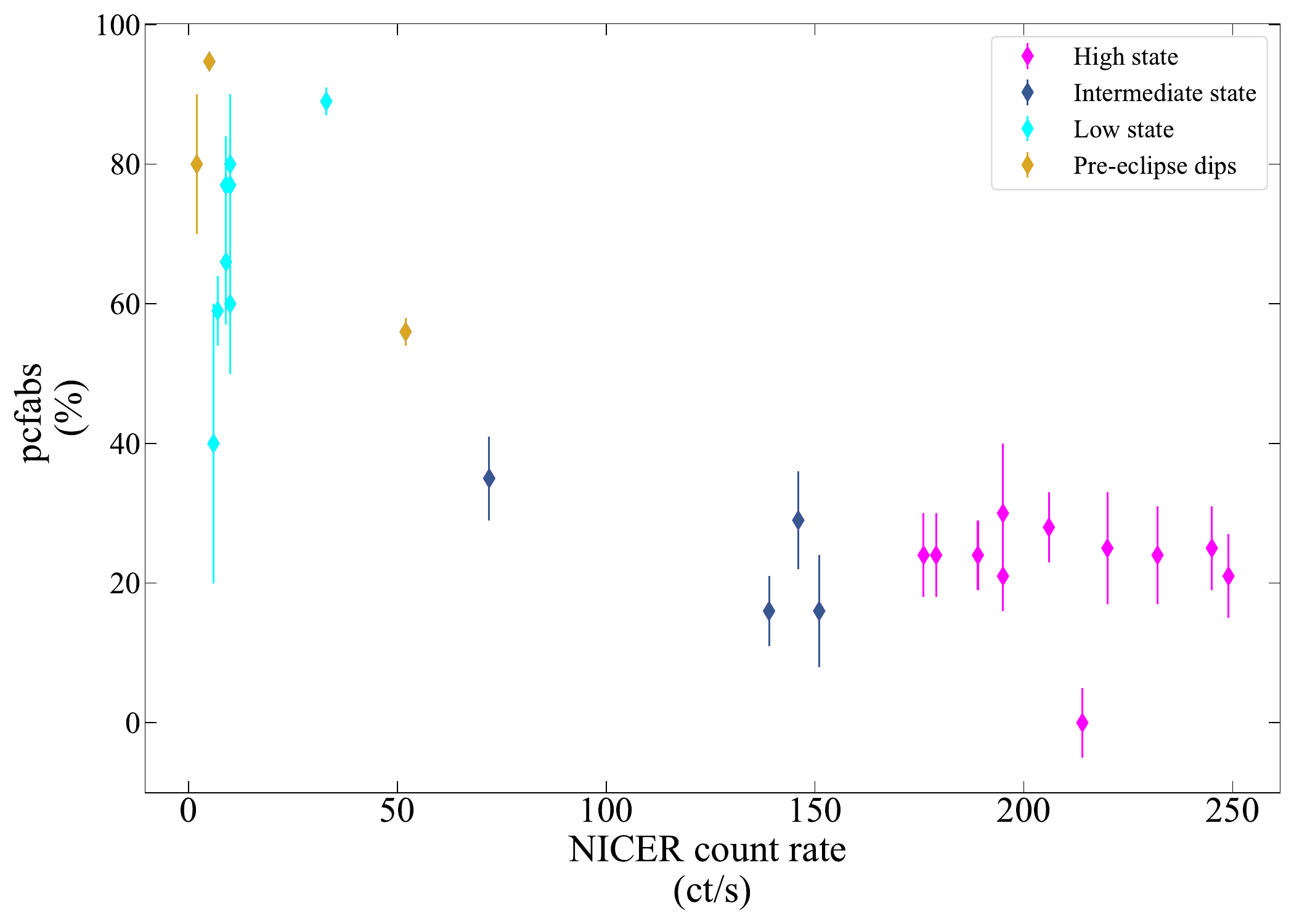}
    \caption{Best fit partial covering fraction versus NICER count rate. The gold points denote pre-eclipse dips, the cyan points are low state observations, the dark blue are intermediate state, and the fuchsia are high state. Overall, the trend between partial covering fraction and superorbital period state appear to be correlated, indicating that the source's low states coincide with a higher percentage of covering, and vice versa. As seen in Table \ref{table:observations}, the observation lengths vary significantly, leading to larger error bars in shorter observations.}
    \label{fig:pcfabs}
\end{figure*}

We implemented a model with an interstellar medium absorption component (\texttt{tbabs}) and a partial covering component (\texttt{pcfabs}) to assess the degree of obscuration from the warped accretion disc \citep[e.g.,][]{Neilsen04}. We set the (\texttt{tbabs}) \nh\ to the average best-fit line of sight value from \cite{Dage22} (2.5$\times 10^{21} \text{cm}^{-2}$), which is broadly consistent with those found in \cite{Neilsen04, Pradhan20} and  \cite{Brumback23}. We freeze this value for all observations, out of concern for degeneracy between the \nh\ of the \texttt{tbabs} and \texttt{pcfabs} models. We also fit a black-body component (\texttt{bbody}), a cut off power-law (\texttt{cutoffpl}) and three Gaussian components for emission lines, with values adopted from \cite{Brumback20}. The $\Gamma$ cut-off value is capped at 20 keV for all observations, similar to \cite{Brumback20} and \citet{Dage22}. Our best-fit model is \texttt{tbabs * pcfabs *(bbody} + cutoffpl + gauss + gauss + gauss).

Except for three pre-eclipse dip observations, the high, intermediate or low states corresponded to the model-predicted count rate, as follows: 
\begin{itemize}
    \item Low,  10 - 50 ct/s
    \item Intermediate, 50 - 150 ct/s
    \item High, $>$150 ct/s
\end{itemize}

We aim to apply the best fit model described above to all observations, regardless of superorbital state, to investigate changes in obscuration or covering fraction due to changes in the accretion disc during excursion. However, spectra from the low states of the superorbital cycle, when the source flux was decreased, had a lower signal-to-noise (S/N) than the high state data sets. Hence, to apply a consistent spectral model to all data sets while reducing degeneracy in the \texttt{pcfabs} components \citep[see][for more discussion on model degeneracy]{Brumback23}, we chose to fix the kT and $\Gamma$ values in the low and intermediate state spectra to the high state average values, which were 0.19 keV and 0.79, respectively. In this way, we are effectively making the assumption that the underlying continuum has a constant spectral shape. This assumption is valid when considering the many historic data sets of this source which consistently produce the same spectral shape of a cut-off power-law and soft blackbody component (e.g. the ASCA observations examined in \citealt{Paul02}). More recent spectral studies of SMC X-1 with NICER over a period of several months also suggest that the underlying spectral shape remains constant \citep{Brumback23}. Assuming a constant spectral shape and freezing the low and intermediate state continuum parameters allows us to test whether this assumption holds for the lower signal to noise data. 

To further assess the impact of freezing the blackbody kT and $\Gamma$ values on our \texttt{pfabs} $N_H$ and covering fraction results, we investigate a representative low state spectrum (ObsID 4509012301). For this fit, we keep the values of kT and $\Gamma$ fixed, but vary their values by up to 20\% from their average values, in increments of 5\%. This was done (1) for kT only, while keeping $\Gamma$ constant, (2) for $\Gamma$ only, while keeping kT constant, and finally (3) while varying both values simultaneously. In all three cases, variations in the values of kT and $\Gamma$ did not result in a change in either \texttt{pcfabs} $N_H$ or covering fraction of more than 5\%. For this reason, we are confident that holding the shape of the spectral continuum constant in the low state data does not induce artificial errors into the \texttt{pcfabs} values. This is consistent with the results found in \citep{Brumback23}, where the spectral continuum in SMC X-1 varies little with superorbital period.   

The best-fit parameters for this model are in Table \ref{table:fits}, and plotted in Figure \ref{fig:allparam}.

\section{Results and Discussion}

\label{sec:res}
The first 26 NICER observations in the MOOSE data set span different stages of the system's superorbital period; that is, its  high, low, and, intermediate stages. In this analysis we closely monitored the variations in the spectral parameters as a function of superorbital period. Our analysis expands upon previous work by \citet{Dage22}, which examined only the high state spectra during excursion and by \citet{Pradhan20}, which did not consider any data taken during excursion.

In Figure \ref{fig:allparam}, the NICER band flux (Panel E) appears to be directly proportional to the flux of the source (Panel F) and its Swift/BAT count rate (Panel G). Since the power law produces the bulk of our continuum spectral model, it is not surprising that it fluctuates in the same way as the flux of the source.

Figure \ref{fig:allparam} demonstrates that certain spectral parameters remain relatively unchanged across different superorbital states. Unsurprisingly, the characteristic disc black body temperature (kT) and power-law photon index ($\Gamma$) of the continuum exhibit minimal variation. These results support previous findings that the shape of SMC X-1's high superorbital state X-ray continuum is stable over time \citep{Dage22,Brumback23}. We expand upon these findings by showing that the low and intermediate state spectra are well fit by a similar continuum model, which supports our assumption that the underlying continuum shape does not change. One possible explanation for the apparent stability of SMC X-1's underlying X-ray continuum is that the inner accretion flow is not sensitive to the rotation of the disc. Such a scenario would imply that the X-ray spectrum does not change with flux state, it simply modulates in normalization and absorption.

In addition to investigating how SMC X-1's excursion X-ray spectrum varied across superorbital state, we also investigated the differences between spectra taken during excursion and those taken after. In Figure \ref{fig:allparam}, the red vertical dashed line marks the end of the excursion period, however there is no abrupt change in the variability of the spectral parameters around this epoch. 

\subsection{Correlation between covering fraction and super-\textbf{orbital} phase}
The \texttt{pcfabs} model parameters \nh \hspace{0.1cm}and covering fraction modulate the shape of the soft end of the X-ray spectrum. Figure \ref{fig:allparam} shows variability in both of these parameters across the time of our observations, although the \nh \hspace{0.1cm}appears to become less variable with time.

Our spectral fitting results indicate a strong negative correlation between \texttt{pcfabs} covering fraction and source count rate, thus implying that the partial covering plays a significant role in the shape of the soft spectrum. To investigate the relationship between covering fraction and super-orbital state, we plotted the covering fraction as a function of the NICER count rate in Figure \ref{fig:pcfabs}. We found that the covering fraction and super-orbital state are inversely related, with the high-state observations having the largest count rates and smallest covering fractions and the low states having the smallest count rates and highest covering fractions. This relationship fits with our understanding of the warped accretion disc precession that drives the super-orbital period.
To investigate the relationship between these two spectral parameters and source count rate in detail, we calculated the Pearson linear correlation coefficient for \nh\ against count rate and for \texttt{pcfabs} against count rate, which provides additional insights.

For the relationship between \nh\ and count rate, we obtained a correlation coefficient of 0.24 with a null hypothesis probability of 0.23, indicating no significant correlation. To determine the influence of outliers, we performed 10$^{4}$ bootstrap simulations. The resulting correlation coefficient distribution revealed a peak at $r=0.25$ with a standard deviation of 0.19, suggesting a large overlap with the null hypothesis. Additionally, we carried out 10$^{4}$ Monte Carlo simulations to test the impact of parameter uncertainties on the correlation coefficient. By considering the \nh\hspace{0.1cm} uncertainty as the standard deviation in a Gaussian distribution for each simulation, we derived a correlation coefficient of $-0.02 \pm 0.2$. This indicates that the correlation between \nh\hspace{0.1cm} and the count rate is statistically negligible.

On the other hand, when examining the anti-correlation between \texttt{pcfabs} and count rate, we found a more substantial negative correlation with a coefficient of $-0.86$, with a null hypothesis probability of $2.3\times10^{-8}$. This implies a consistent inverse relationship between \texttt{pcfabs} and count rate. Employing the bootstrap method resulted in a single-peaked distribution of the correlation coefficient, with an average value of $-0.86$ and a standard deviation of $0.04$. Similarly, Monte Carlo simulations produced a correlation coefficient of $-0.83 \pm 0.04$, consistent with the bootstrap findings. Thus, the observed anticorrelation between \texttt{pcfabs} and the count rate is highly significant compared to any relation between \nh\ and the count rate.
 
Under a model in which absorption by a warped, precessing disc is responsible for the superorbital modulation, we expect maximum obscuration during the super-orbital low states and minimum obscuration during the super-orbital high states. The only places where we see deviation from the inverse relationship between partial covering fraction and super-orbital period are during pre-eclipse dips (see grey dotted lines in Fig. \ref{fig:allparam}). These dips are orbital in nature and therefore can occur during the orbital high state. Our dataset captured 3 pre-eclipse dips, defined as having an orbital phase higher than 0.8, based on the ephemeris from \citet{2015A&A...577A.130F}. These three observations, ObsIDs 4509010401, 4509011201, 4509012301 all showed high covering fractions, which agrees with the interpretation of similar features in Her X-1 (e.g., \citealt{Giacconi73}) as increased obscuration by the ``splash zone'' where the accretion stream impacts the disc. 

Additionally, \citet{Brumback23} analysed 18 NICER observations of SMC X-1 during four different super-orbital high states and found that the continuum parameters (blackbody temperature kT and $\Gamma$) showed very little change between super-orbital high states. Instead, they found that the most significant changes in spectral shape occurred between spectra from the superorbital intermediate state (when the source is increasing in luminosity as the accretion disc moves out of the line of sight) and the high state. \citet{Brumback23} suggested that the underlying spectral continuum shape is constant, but is sensitive to changes in obscuration caused by the rotation angle of the disc. Our results from this analysis, particularly the inverse relationship between the covering fraction and superorbital period state, support these results and the conclusion that the inner accretion flow in SMC X-1 is insensitive to changes in the outer disc shape, although we are assuming that the underlying model continuum does not change significantly in the high/low state. 

While our assumption of a constant spectral state seems well supported by the data, the full picture may be more complex. \cite{Pradhan20} jointly fit SUZAKU and NuSTAR spectra and found that the power-law normalization (in the hard energy band of both detectors) is variable. From their spectroscopic fits, they conclude that there is evidence for a change in the partial covering fraction of the inner disc region, correlated with the superorbital variation. However, due to the power-law variability in the hard X-ray band, they suggest that there is another mechanism behind the superorbital variation, i.e., a change in the instantaneous accretion rate. We cannot make a direct comparison to the \cite{Pradhan20} spectra as our NICER observations do not extend above 12 keV, and do not fully constrain the power-law cut-off (as noted in \citealt{Dage22}, the cut-off extends beyond the range to which NICER is sensitive). Like \citet{Pradhan20}, a correlation between the power-law normalization component ($K$~of {\tt xspec}'s {\tt cutoffpl}) and the 0.3-12.0 keV model flux is apparent in the excursion data (panels E and F of Figure \ref{fig:allparam}), as expected for any major component of the source emission. However, our fits strongly suggest that the only correlation between the superorbital period and the X-ray spectral shape in the soft band is due to variation in the partial covering fraction of the inner acccretion disc.

\begin{figure*}
   \centering
    \includegraphics[scale=0.43]{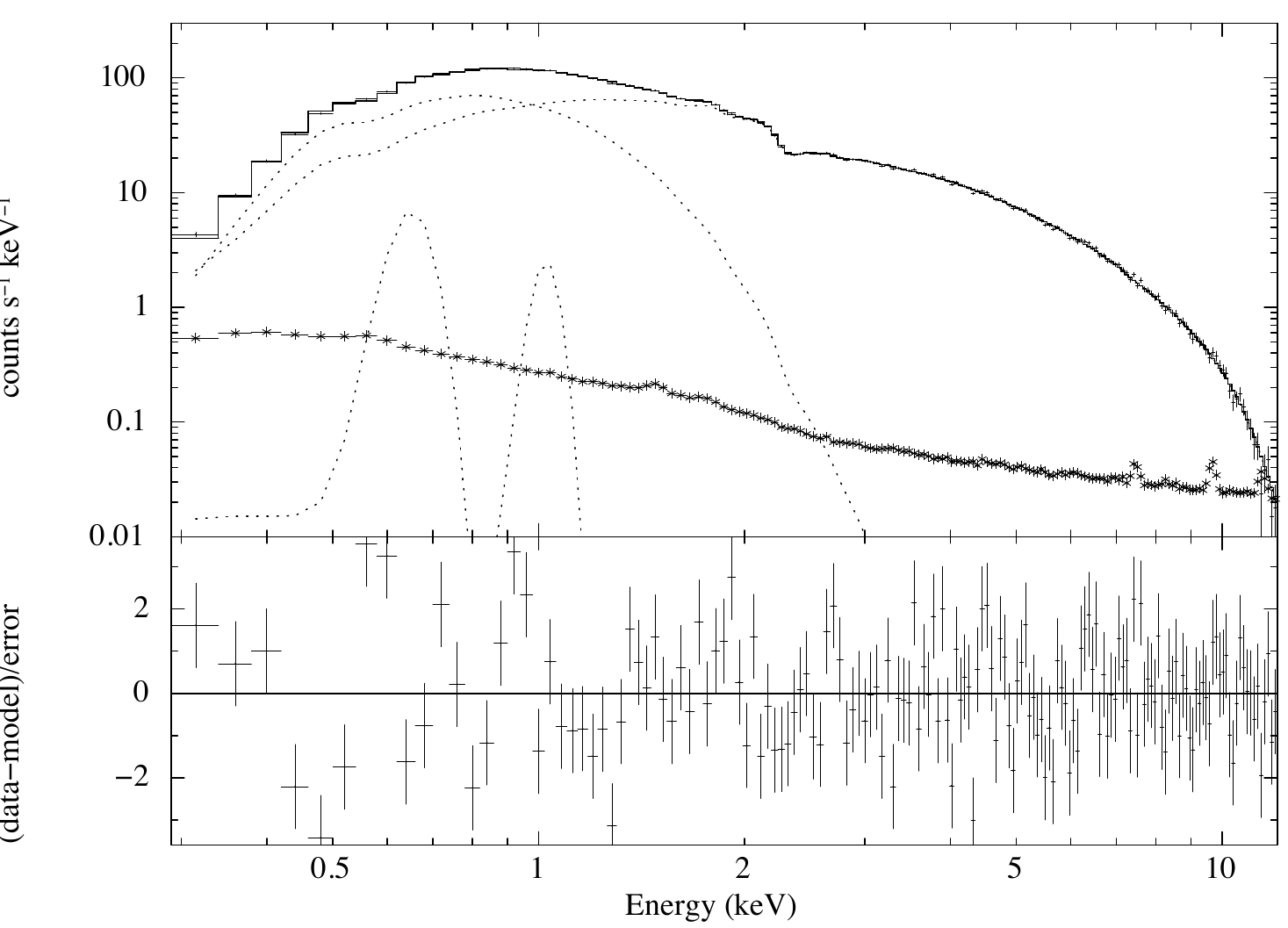}
   \includegraphics[scale=0.43]{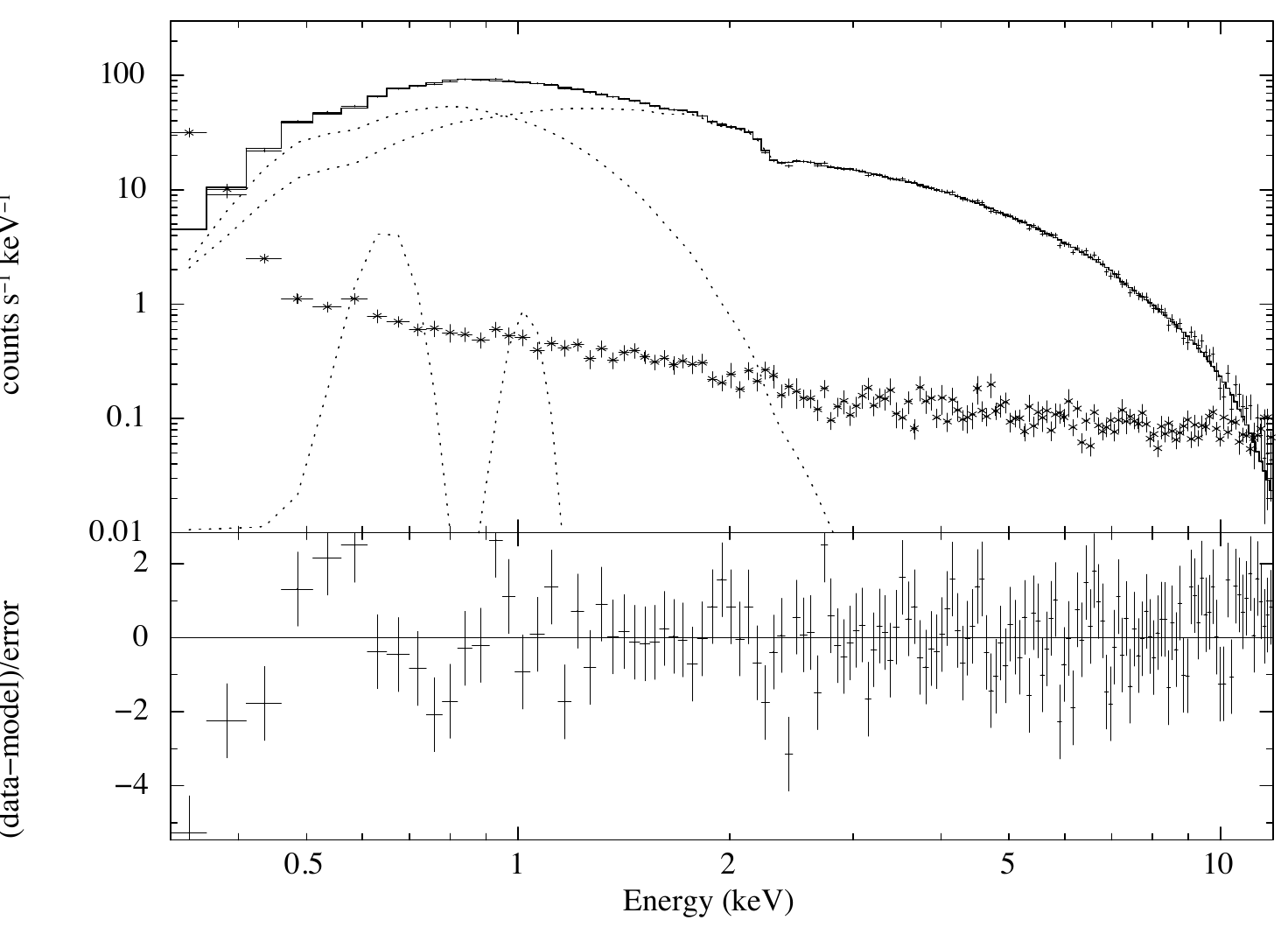}
    \includegraphics[scale=0.43]{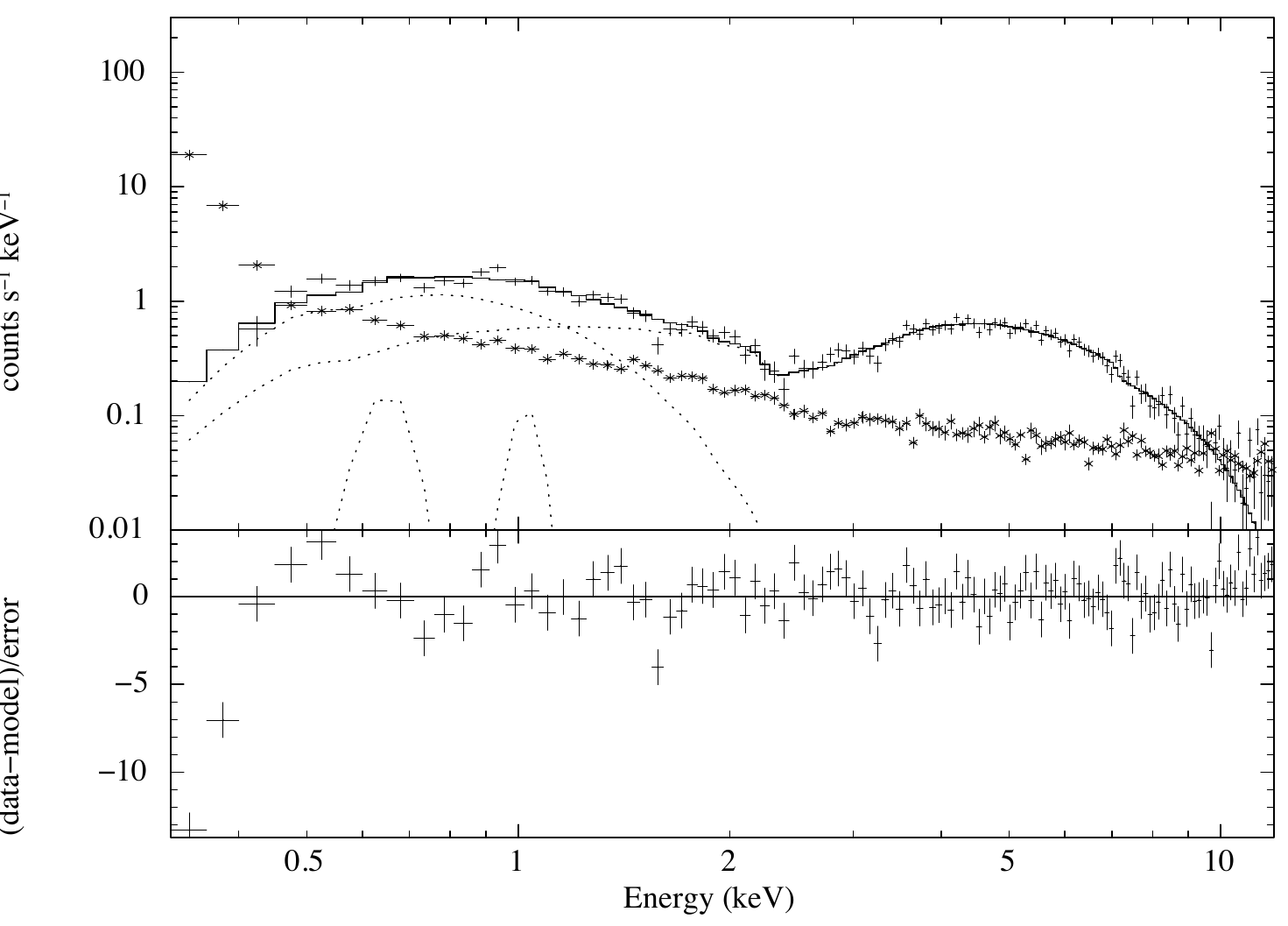} 
   \caption{Representative high state (top panel, ObsID 4509011701), intermediate state (middle panel, ObsID 4509012401), and low state (bottom panel, ObsID 4509012301) spectra fit with our best fit model.}
   \label{fig:high}
\end{figure*}

\section{Conclusions and Future Work}
\label{sec:conc}

We performed X-ray spectroscopy of 26 NICER observations of the neutron star pulsar SMC X-1 to investigate changes in spectral shape across superorbital states within excursion for the first time, and compare to data from outside excursion. Our best fit model for these observations is a black-body and cut-off power-law continuum with several emission lines that is absorbed by a neutral absorber and a partial covering fraction, and we assume that the underlying model continuum does not change significantly in the low state. Our findings are summarized as follows:
\begin{itemize}

    \item The same spectral model can adequately describe all of the data regardless of super-orbital state or excursion status. 
    \item The partial covering parameter is inversely related to source count rate (a proxy for super-orbital state). \citet{Pradhan20} observed this relationship for non-excursion data, and the trend carries on in excursion. This implies a warped accretion disc is obscuring the source, and the super-orbital period is caused by the rotating warped accretion disc covering flux from the neutron star. 
    \item The spectral parameters show no change in the immediate vicinity of the end of excursion. \nh\ potentially shows a transition from an epoch of higher variability to a more stable epoch approximately 70 days before the end of excursion, but the correlation between these behaviors in unclear.

\end{itemize}

The next step will be to study the later observations in the MOOSE data set, which continue after the end of excursion and could provide more insights into the difference in spectral shape and parameter variability between excursion and non-excursion data. 

The presence of soft emission lines in our data is also interesting. These disc-formed emission lines are effective observational tracers of how material in the accretion disc behaves and evolves over time. Throughout a binary orbit, line profile shape/strength will change, carrying within it an imprint of the evolving source of X-rays heating t3he accretion disc itself. An empirical connection between the line emitting regions, and physical properties of the X-ray source heating the disc, has proved to be an effective observational tool for understanding the structure and geometry of the gas making up accretion discs in LMXBs see \citep{tetarenko2020,tetarenko2023}. In the future, a similar technique used for LMXBs could realistically be adapted and applied to SMC X-1, using a suitable model for X-ray irradiation in X-ray pulsar sources (e.g., \citealp{Hickox05}). 

%

\section*{Acknowledgements}

The authors thank the NICER team for flexible scheduling of the observations, and Alex Tetarenko for helpful discussions. RK acknowledges the Trottier Space Institute at McGill summer undergraduate fellowship program for supporting this project. KCD and DH acknowledge funding from the Natural Sciences and Engineering Research Council of Canada (NSERC) and the Canada Research Chairs (CRC) program. KCD acknowledges fellowship funding from Fonds de Recherche du Qu\'ebec $-$ Nature et Technologies, Bourses de recherche postdoctorale B3X no. 319864 and support
provided by NASA through the NASA Hubble Fellowship grant
HST-HF2-51528 awarded by the Space Telescope Science Institute, which is operated by the Association of Universities for Research in Astronomy, Inc., for NASA, under contract NAS5–26555. MCB acknowledges support from NASA Grant 80NSSC23K0619. CPH acknowledges support from the National Science and Technology Council in Taiwan through grant 112-2112-M-018-004-MY3. BET acknowledges support from the Trottier Space Institute (TSI) at McGill, through an TSI Fellowship.

This research has made use of data and/or software provided by the High Energy Astrophysics Science Archive Research Center (HEASARC), which is a service of the Astrophysics Science Division at NASA/GSFC.

\section*{Data Availability}
The NICER observations are publicly available through HEASARC (\url{https://heasarc.gsfc.nasa.gov/docs/archive.html}). 




\bibliographystyle{mnras}
\bibliography{example} 





\bsp 
\label{lastpage}
\end{document}